\documentclass{preprint}

\usepackage{graphicx}

\usepackage{amsmath}
\usepackage{amsfonts}
\usepackage{amssymb}
\usepackage{booktabs}
\usepackage[symbol]{footmisc}
\usepackage{algorithm}
\usepackage{algpseudocode}

\title{A Span-based Model for Extracting Overlapping PICO Entities from RCT Publications}

\author[1]{Gongbo Zhang}
\author[2]{Yiliang Zhou}
\author[3]{Yan Hu}
\author[3]{Hua Xu}
\author[1]{Chunhua Weng$^*$}
\author[2]{Yifan Peng\thanks{Equal contribution. Correspondence to \texttt{cw2384@cumc.columbia.edu}, \texttt{yip4002@med.cornell.edu}.}}
\affil[1]{Department of Biomedical Informatics, Columbia University, New York, New York, US}
\affil[2]{Department of Population Health Sciences, Weill Cornell Medicine, New York, New York, USA}
\affil[3]{Department of Biomedical Informatics \& Data Science, Yale School of Medicine, New Haven, Connecticut, USA}

\begin{document}

\maketitle

\begin{abstract}

\textbf{Objectives:} Extracting PICO (Populations, Interventions, Comparison, and Outcomes) entities is fundamental to evidence retrieval. We present a novel method, PICOX, to extract overlapping PICO entities.

\textbf{Materials and Methods:} PICOX first identifies entities by assessing whether a word marks the beginning or conclusion of an entity. Then, it uses a multi-label classifier to assign one or more PICO labels to a span candidate. PICOX was evaluated using one of the best-performing baselines, EBM-NLP, and three more datasets, i.e., PICO-Corpus and RCT publications on Alzheimer's Disease or COVID-19, using entity-level precision, recall, and F1 scores.

\textbf{Results:} PICOX achieved superior precision, recall, and F1 scores across the board, with the micro F1 score improving from 45.05 to 50.87 (p $\ll$ 0.01). On the PICO-Corpus, PICOX obtained higher recall and F1 scores than the baseline and improved the micro recall score from 56.66 to 67.33. On the COVID-19 dataset, PICOX also outperformed the baseline and improved the micro F1 score from 77.10 to 80.32. On the AD dataset, PICOX demonstrated comparable F1 scores with higher precision when compared to the baseline.

\textbf{Conclusion:} PICOX excels in identifying overlapping entities and consistently surpasses a leading baseline across multiple datasets. Ablation studies reveal that its data augmentation strategy effectively minimizes false positives and improves precision.

\end{abstract}


\section{Background and Significance}
\label{background-and-significance}

The PICO (Population, Intervention, Comparison, and Outcome) framework is a widely adopted standard for evidence retrieval in evidence-based medicine~\cite{Richardson1995-wi, Kang2023-EvidenceMap}. This framework specifies the study population of a clinical study, the intervention and comparison applied, and the expected outcomes to assist with retrieving relevant evidence~\cite{Peng2023-ke}. However, manual extraction of PICO is time-consuming~\cite{Borah2017-dm}. The exponential growth of randomized controlled trial (RCT) publications further exacerbates this challenge. It is imperative to automate PICO extraction to enable timely and efficient evidence retrieval, appraisal, and synthesis.

\begin{figure}
    \centering
    \includegraphics[width=.5\textwidth]{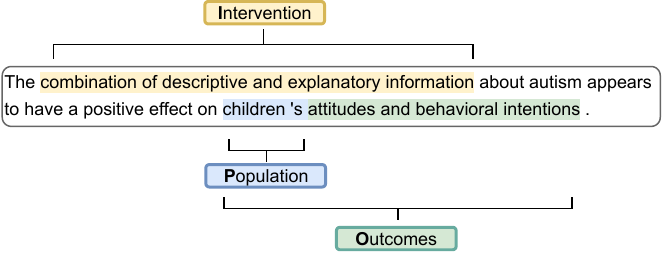}
    \caption{An example sentence that contains overlapping PICO entities. The population entity is contained within the outcome entity.}
    \label{fig:1}
\end{figure}

Existing related studies largely treat PICO extraction as a sequence labeling task of Named Entity Recognition (NER), where each token is labeled with a predefined tag that indicates the entity type paired with the IOB2 schema, symbolizing the inside, outside, and beginning of an entity. Some methods are based on Conditional Random Fields with well-designed features~\cite{McCallum2003-sh, Finkel2005-sf}. However, with the evolution of neural networks, the LSTM-CRF model (Long-Short Term Memory - Conditional Random Field) has demonstrated substantial promise for the NER task~\cite{Ma2016-ru, Yang2018-fn, Kang2019-Pretraining}. More recently, language models such as ELMo~\cite{Peters2018-jq} and BERT~\cite{Devlin2019-vb} have also achieved remarkable successes~\cite{Gu2021-gu, Wang2022-vz, Beltagy2019-kx, Lee2020-qp,Kanakarajan2021-lh}. While these methods have demonstrated promising results, they often falter when handling overlapping PICO entities, which are commonly observed in PICO annotations and account for 8.2\% of the sentences in the EBM-NLP dataset~\cite{Nye2018-jy}. For example, in Figure \ref{fig:1}, in the text of ``\emph{children's attitude and behavior interventions},'' there are overlapping P entity (``children'') and O entities (``children's \emph{attitude''} and \emph{``(children's) behavior intentions}'').

Span-based methods have been frequently adopted to extract overlapping entities~\cite{Wadden2019-wx, Luan2019-ay, Liu2020-sl, Tan2020-oq, Fu2021-ee, Li2021-ew, Wan2022-ij, Zhu2022-ki}. These methods identify spans containing named entities and classify these candidates by corresponding entity types but have not yet been used for PICO extraction. Moreover, to limit the number of span candidates, most studies presume a maximum length for an entity span~\cite{Golam_Sohrab2021-fm, Fei2021-av, Zaratiana2022-sx, Zaratiana2022-xb}. While this presumption works effectively for generic named entities like people or locations, it is not necessarily optimal for PICO entity extraction. Regarding PICO entities, the length of entity spans has a significant variability. Some can be as condensed as a single word, like treatments, while others can expand across an entire sentence, providing comprehensive descriptions of population groups, including age, gender, and sample size. Hence, assuming a maximum length for an entity span might fail to capture PICO entities of various lengths.

\begin{figure}
    \centering
    \includegraphics[width=.8\textwidth]{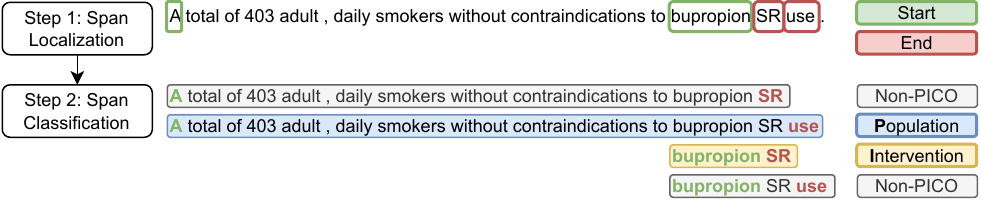}
    \caption{The workflow for PICOX consists of two steps. Initially, it detects the start and end positions in the text sequence. Subsequently, it categorizes each span bounded by a pair of start and end positions. In the provided example from PMID 20840173, PICOX discerned two potential start positions and two end positions, resulting in four valid pairs where the start position does not surpass the end position. Post span classification, PICOX identified a population entity "A total of 403 adult ... to bupropion use" and an intervention entity "bupropion SR". In this instance, the Intervention entity is encapsulated within the Population entity.}
    \label{fig:2}
\end{figure}

To tackle these unsolved problems, we define PICO extraction as a ``span detection'' task and present a novel span-based model called PICOX. Figure \ref{fig:2} shows the two-step workflow for PICOX. At step 1, drawing inspiration from the works of Tan et al.~\cite{Tan2020-oq} and Shen et al.~\cite{Shen2021-jp}, it locates the entities in a sentence by determining whether a word signals the start or end of an entity. It is distinguished from previous works by not only assessing if a token marks the start or end of spans but also categorizing non-boundary tokens as ``inside'' or ``outside'' spans. This distinction provides advantages for model training, as it considers the relative orders of these tokens. For instance, the likelihood of the last token in an entity being followed by a word outside any entity is higher than by a word inside one. At step 2, it classifies each span using a multi-label classifier. The objective here is to not only identify the entity type of a span, but also to differentiate spans that represent an entity from those that do not. Since a sentence might contain multiple entities, not all extracted span candidates represent a valid entity; refer to the 1st and 4th extracted spans in Figure \ref{fig:2} for examples. To distinguish PICO entities and such invalid span candidates, we introduce a new strategy that augments the training data.

This research makes three key contributions. First, we contribute a novel framework that parses overlapping spans and accommodates unlimited lengths of entities. Secondly, we present a data augmentation strategy to boost the training data for the span classifier, which not only identifies the entity type within a span but also differentiates between spans that represent an entity and those that do not. Thirdly, we extensively evaluate our model on four diverse benchmark datasets.

\section{Materials and Methods}
\label{materials-and-methods}

\subsection{Data description}
\label{data-description}

This study utilized four benchmark datasets: EBM-NLP~\cite{Nye2018-jy}, PICO-Corpus~\cite{Mutinda2022-nm}, and two sets of RCT publications focusing on Alzheimer's Disease and COVID-19 (Table \ref{tab:1})~\cite{Hu2023-qh}.

\begin{table}
\caption{Number of annotations in the datasets. P - Population. I - Intervention/Comparison. O - Outcome. AD - Alzheimer's Disease.}\label{tab:1}
\centering
\small
\begin{tabular}{lrrrrrrrrrr}
\toprule
Dataset & \multicolumn{3}{c}{Training} & \multicolumn{3}{c}{Test} & \multicolumn{3}{c}{Total} & \% of overlap 
\\
\cmidrule(rl){2-4}\cmidrule(rl){5-7}\cmidrule(rl){8-10}
& P & I & O & P & I & O & P & I & O & entities\\
\midrule
EBM-NLP & 17,952 & 32,859 & 33,554 & 643 & 1,726 & 1,833 & 18,595 & 34,585 & 35,387 & 8.2\% \\
PICO-Corpus & 1,404 & 1,849 & 3,580 & 1,561 & 2,054 & 3,978 & 3,121 & 4,108 & 7,956 & 0\% \\
AD & 98 & 221 & 295 & 109 & 245 & 328 & 218 & 490 & 656 & 0\% \\
COVID-19 & 118 & 281 & 279 & 132 & 313 & 310 & 263 & 625 & 619 & 0\% \\
\bottomrule
\end{tabular}
\end{table}

\subsubsection{EBM-NLP}
\label{ebm-nlp}

The EBM-NLP dataset consists of 4,993 abstracts describing CRT~\cite{Nye2018-jy}. These abstracts were annotated with entities related to ``population'', ``intervention/comparison'', and ``outcome''. The annotations for interventions and comparisons were merged during the annotation process. The training labels for this dataset were obtained via crowdsourcing on Amazon Mechanical Turk and subsequently aggregated. The test set, in contrast, was manually labeled by medical professionals. We used the standard training and test sets provided in the EBM-NLP. Following a previous practice~\cite{Zhang2018-zc}, we randomly selected 5\% from the training set for evaluation.

\subsubsection{Alzheimer's Disease (AD) and COVID-19 datasets}
\label{alzheimers-disease-ad-and-covid-19-datasets}

We collected a total of 1,980 abstracts from RCT publications related to Alzheimer's disease (AD) (as of November 5, 2021) and 552 abstracts related to COVID-19 (as of September 1, 2021). From these collections, we randomly selected and annotated 150 RCT abstracts for each disease~\cite{Hu2022-sb}. To enhance the annotation process, we used a sentence classification model that filtered and focused solely on method-related sentences. This narrowed the scope of annotation~\cite{Hu2023-qh}. The RCT abstracts filtered through the sentence classification model were annotated by two independent annotators, both with medical training. To ensure high accuracy and consistency, we followed a rigorous annotation test loop grounded in the original EBM-NLP annotation guidelines. These were supplemented with rules and examples in each iteration to improve clarity.

In each loop, ten abstracts were selected randomly and manually annotated using the evolving guidelines. After each loop, we evaluated the inter-annotator agreement using Cohen's kappa statistic. If the measure of agreement was below 0.7, we addressed the discrepancies between annotators and refined the guidelines by enriching rules and examples. By conducting multiple annotation loops and continuously improving the guidelines, we achieved a final inter-annotator agreement of 0.714, 0.808, 0.701, and 0.790 for Population, Intervention, Comparison, and Outcome, respectively. The combined coefficient for all PICO elements stood at 0.746~\cite{Hu2023-qh}.

With the finalized guidelines, we annotated the above abstracts in this study. To create a sufficiently large test set, we divided the PICO-Corpus into two subsets of the same size, one for training and the other for testing. Following an earlier biomedical text mining work, we take 5\% of samples out of the training set as the validation set, to tune the model's hyperparameters~\cite{Zhang2018-zc}.

\subsubsection{PICO-Corpus}
\label{pico-corpus}

The PICO-Corpus comprises 1,011 PubMed abstracts, all of which are RCTs primarily focusing on breast cancer~\cite{Mutinda2022-nm}. Each abstract was annotated with specific textual elements that represent the Participants, Intervention, Control, and Outcome (PICO). Like the EBM-NLP dataset~\cite{Nye2018-jy}, the annotations for interventions and comparisons were consolidated into one category. Like the AD and COVID-19 corpus, we divided the PICO-corpus datasets into three distinct subsets: 45\% for training, 5\% for validation, and the rest 50\% for testing.

\subsection{Proposed model}
\label{proposed-model}

PICOX comprises two parts (Figure \ref{fig:2}): one to identify the presence of entity spans in a sentence and the other to determine the types of the located span.

\subsubsection{Span localization}
\label{span-localization}

The input of the model is a sequence of text with length $n$, denoted as $s = \{ t_{1},t_{2},...,t_{n}\}$. Then, we apply a BERT-based encoder to obtain contextual representations of each token, denoted as $H = \{ h_{1},h_{2},...,h_{n}\}$. Next, we develop a model to determine whether a token marks the beginning or end of an entity. In this case, we define a set of five relative-position categories $L =$ \{\emph{inside}, \emph{outside}, \emph{start}, \emph{end}, \emph{both-start-and-end}\}. The ``both-start-and-end'' category is incorporated for situations where a single-word entity results in a token functioning as both the start and end of a span. Because a sentence may contain multiple entities, we identify the start and end of all entities present in the sentence. We calculate the probability $P_{ij}^{L}$ of the token $t_{i}$ being in relative position $j \in L$ to the nearest entity.

\begin{equation}
P_{ij}^{L} = \frac{\exp\left( W_{j}^{L} \cdot h_{i} + b_{j}^{L} \right)}{\sum_{l \in L}^{\ }\ \exp\left( W_{l}^{L} \cdot h_{i} + b_{l}^{L} \right)}
\end{equation}

where $W^{L}$ and $b^{L}$ are parameters to be learned. We use the cross-entropy as the loss function,

\begin{equation}
L^{L} = \sum_{i = 1}^{N}\ \sum_{j \in L}^{\ }\  - y_{ij}^{L} \cdot logP_{ij}^{L}
\end{equation}

where $y_{ij}^{L}$ denotes the gold standard relative-position label of token $t_{i}$.

To determine whether a token $t_{i}$ is the starting point of a span, we select a set of relative positions $L_{S} = \{ j|{\hat{P}}_{ij}^{L} \geq t\}$, where $t$ represents a threshold parameter. We then consider position $j$ as a starting point if either ``both-start-and-end'' or ``start'' is within the set $L_{S}$. We identify ending positions using the same procedure.

\subsubsection{Span classification}
\label{span-classification}

After predicting the span boundaries, we extract all spans $t_{s:e}$ defined by pairs of start and end positions $s$ and $e$ where $s < e$. Subsequently, we classify these spans based on the type of entity, denoted as $C =$ \{Population, Intervention/Comparison, Outcome\}. Following the work of Nye et al~\cite{Nye2018-jy}, we merge Interventions and Comparison into a single category (I). For each span $t_{s:e}$, we feed $t_{s:e}$ into BERT, adding a special token {[}CLS{]} at the beginning, and utilize the encoding of the inserted token {[}CLS{]}, denoted as $h_{s:e}$, as the feature vector~\cite{Fei2021-av}. To determine whether span $t_{s:e}$ falls under category $c \in C$, we calculate

\begin{equation}
P_{s:e}^{c} = \sigma\left( W^{c} \cdot h_{s:e} + b^{c} \right)
\end{equation}

where $\sigma( \cdot )$ is the Sigmoid function and $W^{c}$ and $b^{c}$ are weight parameters. The objective function of span classification is defined as

\begin{equation}
L^{C} = \sum_{c \in C}^{\ }\  - y_{s:e}^{c} \cdot logP_{s:e}^{c}
\end{equation}

where $y_{s:e}^{c}$ indicates whether the span $t_{s:e}$ represents an entity span in the category $c$.

\subsection{Data augmentation}
\label{data-augmentation}

In scenarios where a sentence contains multiple entities, some extracted spans may not represent any entity. For example, consider a sentence with two spans, $t_{s1}...t_{e1}$ and $t_{s2}...t_{e2}$, containing two pairs of start and end positions. This would result in a total of four span candidates being extracted. However, two of these candidates, $t_{s1}...t_{e2}$ and $t_{s2}...t_{e1}$, are not associated with any PICO elements. We refer to such candidates as ``\emph{composite spans''} because their boundaries are made up of the start of one entity and the end of another.

Throughout the span classification stage, it is necessary to filter out the composite spans from the extracted candidates and accurately identify the entity type for the remaining ones. To address this challenge, we augment span classification training data by including composite spans (Algorithm \ref{alg:augment_data}). By introducing these extra spans, we provide the model with examples that aid it in learning to differentiate between spans representing named entities and those that do not.

\vspace{1em}

\begin{algorithm}[t]
\caption{Data augmentation.}\label{alg:augment_data}
Let $l_a$, $l_b$ be two entity lists of different categories.
\begin{algorithmic}

\State $D \leftarrow \{\}$
\For {($e_a$, $e_b$) in $l_a \times l_b$}:
\If{START($e_a$) $\leq$ END($e_b$)}
    \State $D \leftarrow D \cup \{<\text{START($e_a$)}, \text{END($e_b$)}>\}$
\EndIf
\If{START($e_b$) $\leq$ END($e_a$)}
    \State $D \leftarrow D \cup \{<\text{START($e_b$)}, \text{END($e_a$)}>\}$
\EndIf
\EndFor
\State \textbf{return} $D$
\end{algorithmic}
\end{algorithm}

\subsection{ Evaluation metrics}
\label{evaluation-metrics}

We employed the span-level precision, recall, and F1 scores as evaluation metrics. For a prediction to be considered a true positive, it must satisfy two conditions: (1) the predicted span is identical to the truth entity, and (2) they should have the same PICO entity type.

\subsection{Experimental settings}
\label{experimental-settings}

In our experiments, we compared PICOX with multiple open-source models, including BioELECTRA~\cite{Kanakarajan2021-lh}, BioBERT~\cite{Lee2020-qp}, SciBERT~\cite{Beltagy2019-kx}, and PubMedBERT~\cite{Gu2021-gu} on EBM-NLP, the largest dataset with PICO annotation. We selected PubMedBERT as our backbone model. To the best of our knowledge, PubMedBERT is one of the best language models for biomedical text. We fine-tuned an instance of the large, uncased version on EBM-NLP. As a sanity check, our baseline instance achieved a token-level macro F1 score of 73.33, closely matching the reported 73.38 in the work of Gu et al.~\cite{Gu2021-gu}.

On the EBM-NLP dataset, we fine-tuned the span localization and span classification models using their respective training data. On the PICO-Corpus, AD, and COVID-19 datasets, we continued fine-tuning the models to benefit from the knowledge learned from the larger EBM-NLP dataset.

We applied a learning rate of 5e-5, batch size 8, and 3 epochs of training. The default span localization threshold is 0.25 (0.4 was used for AD). Intel Core i9-9960X 16 cores processor, NVIDIA Quadro RTX 5000 GPU, and a memory size of 128G were used in this work.

\section{Results}
\label{results}

\subsection{ Overall performance}
\label{overall-performance}

We evaluated PICOX on the following benchmark datasets: EBM-NLP~\cite{Nye2018-jy}, PICO-Corpus~\cite{Mutinda2022-nm}, Alzheimer's Disease, and COVID-19 trials~\cite{Hu2023-qh}. The results demonstrate that our model outperforms the previously established state-of-the-art BERT-based models~\cite{Gu2021-gu, Beltagy2019-kx, Lee2020-qp, Kanakarajan2021-lh}. Our model significantly (p $\ll$ 0.01) enhances the identification of both overlapping entities (with a 4.82\% increase in F1 score) and non-overlapping entities (with a 5.16\% increase in F1 score). Further analysis shows that the data augmentation strategy effectively reduces false positive errors which result in a higher precision.

Table \ref{tab:2} compares PICOX with BioELECTRA, BioBERT, SciBERT, PubMedBERT on the EBM-NLP. In this study, we selected PubMedBERT as the baseline since it was pretrained specifically on biomedical text corpus and achieved a similar or better overall performance (Macro F1) than others for PICO extraction~\cite{Gu2021-gu} (Table \ref{tab:2}). We replicated the performance of the baseline model, as reported by Gu et al.~\cite{Gu2021-gu}. In the paired t-test, our method significantly outperformed PubMedBERT with a p-value $\ll$ 0.01.

\begin{table}[!tbp]
\caption{Performance comparison of BioELECTRA, SciBERT, BioBERT, PubMedBERT, and PICOX on EBM-NLP. P - Precision, R - Recall. F1 - F1 score.}\label{tab:2}
\centering
\footnotesize
\begin{tabular}{lccccccccccc}
\toprule
& \multicolumn{3}{c}{\textbf{Participants}} & \multicolumn{3}{c}{\textbf{Interventions}} & \multicolumn{3}{c}{\textbf{Outcomes}} & \multicolumn{2}{c}{\textbf{Average F1}} \\
\cmidrule(rl){2-4}\cmidrule(rl){5-7}\cmidrule(rl){8-10}\cmidrule(rl){11-12}
& \textbf{P} & \textbf{R} & \textbf{F1} & \textbf{P} & \textbf{R} & \textbf{F1} & \textbf{P} & \textbf{R} & \textbf{F1} & \emph{micro} & \emph{macro} \\
\midrule
BioELECTRA & 52.30 & 63.76 & 57.46 & 52.86 & 41.83 & 46.70 & 44.40 & 33.93 & 38.47 & 45.25 & 47.54 \\
SciBERT & 55.36 & 62.67 & 58.79 & 51.77 & 41.43 & 46.03 & 45.07 & 33.17 & 38.21 & 45.07 & 47.68 \\
BioBERT & 54.19 & 62.36 & 57.99 & 55.02 & 41.89 & 47.57 & 44.69 & 33.06 & 38.01 & 45.45 & 47.85 \\
PubMedBERT & \textbf{56.38} & 63.91 & 59.91 & 51.32 & 41.54 & 45.92 & 44.44 & 32.90 & 37.81 & 45.05 & 47.88 \\
PICOX & 55.93 & \textbf{66.72} & \textbf{60.85} & \textbf{57.13} & \textbf{52.43} & \textbf{54.68} & \textbf{48.25} & \textbf{38.41} & \textbf{42.77} & \textbf{50.87} & \textbf{52.11} \\
\bottomrule
\end{tabular}
\end{table}

To demonstrate our method's generalizability, we also compared the performance of PICOX on the EBM-NLP, PICO-Corpus, AD, and COVID-19 datasets with PubMedBERT, as shown in Table \ref{tab:3}. On the EBM-NLP dataset, PICOX achieved higher precision, recall, and F1 scores across all categories than the baseline, with the micro F1 score increasing from 45.05 to 50.87. Specifically for interventions, the precision is improved from 51.32 to 57.13, the recall is improved from 41.54 to 52.43, and F1 is improved from 45.92 to 54.68.

On the PICO-Corpus, PICOX achieved lower precision but higher recall and F1 scores than the baseline. The micro recall score improves from 56.66 to 67.33, and the micro F1 score increases from 62.87 to 65.64. For the interventions, the recall was increased from 51.68 to 65.24. The result underscores PICOX's capability to correctly classify entities within the PICO-Corpus.

On the AD dataset, PICOX demonstrated comparable F1 scores with higher precision but lower recall when compared to the baseline. This result suggests that PICOX achieved more accurate predictions but misclassified certain entities as non-PICO elements. Nevertheless, the overall F1 score remained similar.

On the COVID-19 dataset, PICOX outperformed the baseline, achieving higher precision, recall, and F1 scores. The micro F1 score was improved from 77.10 to 80.32, indicating the effectiveness of PICOX in accurately extracting and classifying entities on the COVID-19 dataset.

\begin{table}
\caption{Performance for PICO entity recognition on EBM-NLP, PICO-Corpus, AD, and COVID-19 datasets. P - Precision, R - Recall. F1 - F1 score.}\label{tab:3}
\centering
\footnotesize
\begin{tabular}{lccccccccccccccc}
\toprule
& \multicolumn{3}{c}{\textbf{EBM-NLP}} &\multicolumn{3}{c}{\textbf{PICO-Corpus}}& \multicolumn{3}{c}{\textbf{AD}} &\multicolumn{3}{c}{\textbf{COVID-19}} \\ 
\cmidrule(lr){2-4} \cmidrule(lr){5-7} \cmidrule(lr){8-10} \cmidrule(lr){11-13}
& \textbf{P} & \textbf{R} & \textbf{F1} & \textbf{P} & \textbf{R} & \textbf{F1} & \textbf{P} & \textbf{R} & \textbf{F1} & \textbf{P} & \textbf{R} & \textbf{F1}\\
\midrule
\multicolumn{2}{l}{PubMedBERT} \\
\hspace{1em}Populations & 56.38 & 63.92 & 59.91 & 69.53 & 78.07 & 73.55 & 83.62 & 80.83 & 82.20 & 82.39 & 87.97 & 85.09 \\
\hspace{1em}Interventions & 51.32 & 41.54 & 45.92 & 67.78 & 51.68 & 58.65 & 71.43 & 74.14 & 72.76 & 73.31 & 73.31 & 73.31 \\
\hspace{1em}Outcomes & 44.44 & 32.90 & 37.81 & 73.99 & 53.85 & 62.33 & 74.01 & 74.92 & 74.46 & 74.32 & 80.95 & 77.49 \\
\hspace{1em}\emph{micro} & 49.70 & 41.19 & 45.05 & 70.61 & 56.66 & 62.87 & 74.58 & 75.64 & 75.11 & 75.30 & 78.99 & 77.10 \\
\hspace{1em}\emph{macro} & 50.71 & 46.12 & 47.88 & \textbf{70.43} & 61.20 & 64.84 & 76.35 & \textbf{76.63} & 76.48 & 76.67 & 80.75 & 78.63 \\
\midrule
\multicolumn{2}{l}{PICOX} \\
\hspace{1em}Populations & 55.93 & 66.72 & 60.85 & 55.92 & 80.09 & 65.86 & 88.29 & 81.67 & 84.85 & 90.16 & 82.71 & 86.27 \\
\hspace{1em}Interventions & 57.13 & 52.43 & 54.68 & 68.12 & 65.24 & 66.65 & 76.69 & 68.82 & 72.55 & 78.98 & 76.07 & 77.50 \\
\hspace{1em}Outcomes & 48.25 & 38.41 & 42.77 & 64.29 & 64.81 & 64.55 & 76.14 & 72.14 & 74.09 & 76.53 & 85.42 & 80.73 \\
\hspace{1em}\emph{micro} & 53.49 & 48.50 & 50.87 & 63.98 & 67.33 & 65.61 & 78.41 & 72.52 & 75.35 & 79.53 & 81.13 & 80.32 \\
\hspace{1em}\emph{macro} & \textbf{53.77} & \textbf{52.52} & \textbf{52.77} & 62.78 & \textbf{70.05} & \textbf{65.69} & \textbf{80.38} & 74.21 & \textbf{77.16} & \textbf{81.89} & \textbf{81.40} & \textbf{81.50} \\
\bottomrule
\end{tabular}
\end{table}

\subsection{Performance on overlapping PICO entities}
\label{performance-on-overlapping-pico-entities.}

To further investigate the discrepancy in performance, we divided the sentences in the EBM-NLP test set into two groups: one containing sentences with overlapping PICO elements and the other without any overlapping entities. We compared the precision, recall, and F1 scores (micro-averaged) across all categories. Table \ref{tab:result_overlap} shows that PICOX consistently outperforms the baseline, with F1 scores improving from 28.15 to 32.97 for the overlapping entity detection and from 46.51 to 51.67 for the non-overlapping entity detection. It is also worth noting that both PICOX and the baseline exhibit lower performance in detecting overlapping entities than non-overlapping ones, highlighting the challenge of identifying overlapping spans in the PICO extraction tasks. Appendix Table \ref{app tab:1} breaks down the over performance on EBM-NLP by the entity length into categories of 1 word, 2-5 words, and \textgreater{} 5 word.

\begin{table}
\caption{Performance of detecting overlapped and non-overlapped entities on EBM-NLP. 
P - Precision. R - Recall. F1 - F1 score.}\label{tab:result_overlap}
\centering
\begin{tabular}{llccc}
\toprule 
&& \textbf{P} & \textbf{R} & \textbf{F1} \\ \midrule
\multicolumn{2}{l}{PubMedBERT} \\ 
& Overlapped & 38.05 & 22.34 & 28.15 \\
 & Non-overlapped & 50.51 & 43.10 & 46.51 \\ \midrule
\multicolumn{2}{l}{PICOX}\\
& Overlapped & 41.09 & 27.53 & 32.97 \\
 & Non-overlapped & 54.22 & 49.36 & 51.67 \\ \bottomrule
\end{tabular}
\end{table}

\subsection{The impact of data augmentation}
\label{the-impact-of-data-augmentation}

To assess the effectiveness of the data augmentation strategy, we compared two variants of our model. In the first variant, we trained the span classifiers on a collection of text spans, each representing a PICO entity. In the second variant, we augmented the training data according to Algorithm 1. We evaluated the performance of the two versions on EBM-NLP and displayed the results in Table \ref{tab:evaluate_span_sampling}. Version 2 achieved higher precision (53.77 vs. 51.49) and maintained a similar recall (52.52 vs. 52.77) compared to version 1.

\begin{table}
\caption{Comparison of two different implementations, one incorporates the data augmentation strategy whereas the other does not. The performance was evaluated on EBM-NLP. P - Precision. R - Recall. F1 - F1 score.}\label{tab:evaluate_span_sampling}
\centering
\begin{tabular}{llccc}
\toprule 
&  & \textbf{P} & \textbf{R} & \textbf{F1} \\ \midrule
\multicolumn{2}{l}{\textit{w/o data augmentation}}\\
& Populations & 50.00 & 66.87 & 57.22 \\
& Interventions & 56.96 & 52.61 & 54.70 \\
& Outcomes & 47.50 & 38.84 & 42.74 \\
& \hspace{1em}\textit{micro} & 51.86 & 48.79 & 50.28 \\
& \hspace{1em}\textit{macro} & 51.49 & \textbf{52.77} & 51.55 \\ \midrule
\multicolumn{2}{l}{\textit{w/ data augmentation}} \\ 
& Populations & 55.93 & 66.72 & 60.85 \\
& Interventions & 57.13 & 52.43 & 54.68 \\
& Outcomes & 48.25 & 38.41 & 42.77 \\
& \hspace{1em}\textit{micro} & 53.49 & 48.50 & 50.87 \\
& \hspace{1em}\textit{macro} & \textbf{53.77} & 52.52 & \textbf{52.77} \\ \bottomrule
\end{tabular}
\end{table}

\subsection{The effectiveness of the span localization threshold}
\label{the-effectiveness-of-the-span-localization-threshold}

In our approach, we utilized a threshold to determine if a word should be considered as the start or end point of an entity span. In this section, we analyzed the effectiveness of this threshold on the model's performance. We plotted the changes in precision, recall, and F1 on the EBM-NLP dataset for varying threshold values, ranging from 0.2 to 0.5. A minimum 0.2 threshold was chosen because we pre-defined five relative-position categories. A maximum 0.5 threshold was chosen because when the threshold is larger than 0.5, the localization model will assign only a single label to each word. Consequently, it may fail to locate spans that consist of a single word, which is both the start and end of the span.

\begin{figure}
    \centering
    \includegraphics[width=\textwidth]{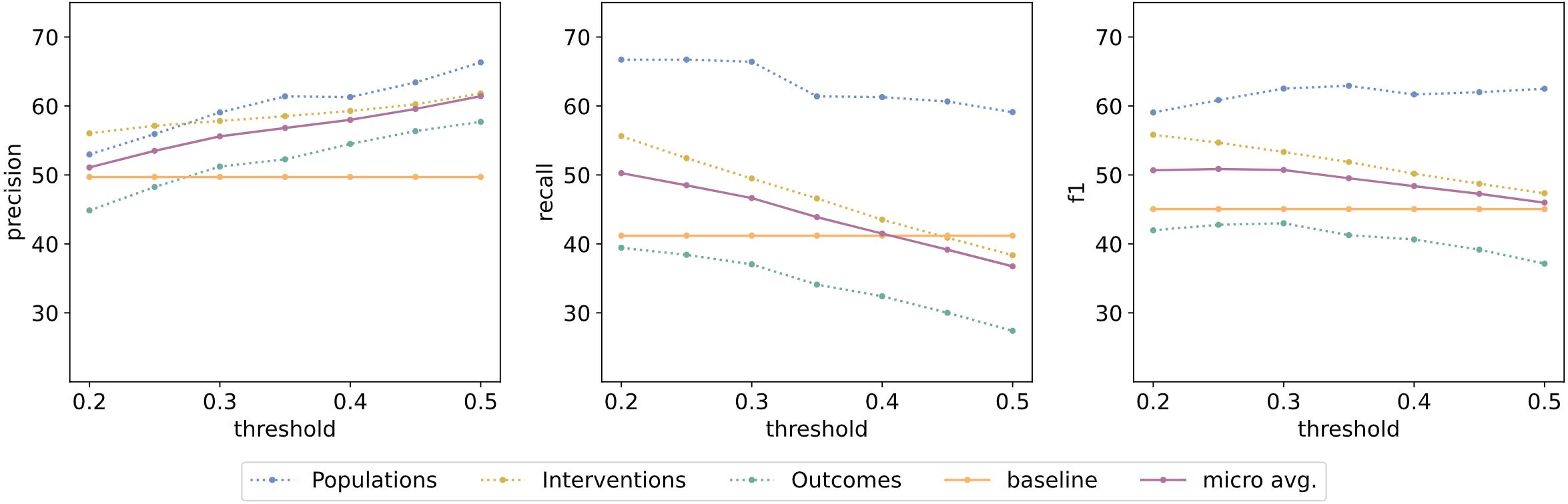}
    \caption{Comparison of two different implementations, one incorporates the data augmentation strategy whereas the other does not. The performance was evaluated on EBM-NLP. P - Precision. R - Recall. F1 - F1 score.}
    \label{fig:3}
\end{figure}

As shown in Figure \ref{fig:3}, an increase in the threshold leads to higher precision but lower recall. This observation aligns with the understanding that a more selective model filters out false positives but may also overlook true positives.

\section{Discussion}
\label{discussion}

Across all four datasets, PICOX achieved superior performance than PubMedBERT. For the EBM-NLP dataset, PICOX surpassed the baseline in precision, recall, and F1 scores, with the micro F1 score rising from 45.05 to 50.87. For the PICO-Corpus, PICOX had similar precision but better recall and F1 scores than the baseline. Notably, the micro recall score went from 56.66 to 67.33. For the AD dataset, PICOX had a similar F1 score and higher precision but lower recall than the baseline. For the COVID-19 dataset, PICOX excelled over the baseline in all metrics.

As shown in Table \ref{tab:result_overlap}, F1 scores rose from 28.15 to 32.97 for overlapping entity detection and from 46.51 to 51.67 for non-overlapping entity detection. Both the PICOX and the baseline had more difficulty detecting overlapping entities than non-overlapping ones, underlining the intricacy of pinpointing overlapping sections in PICO extraction tasks. These results show that PICOX performs better than the baseline at identifying entities that are contained within another longer one, i.e., overlapping spans.

\begin{figure}
    \centering
    \includegraphics[width=.8\textwidth]{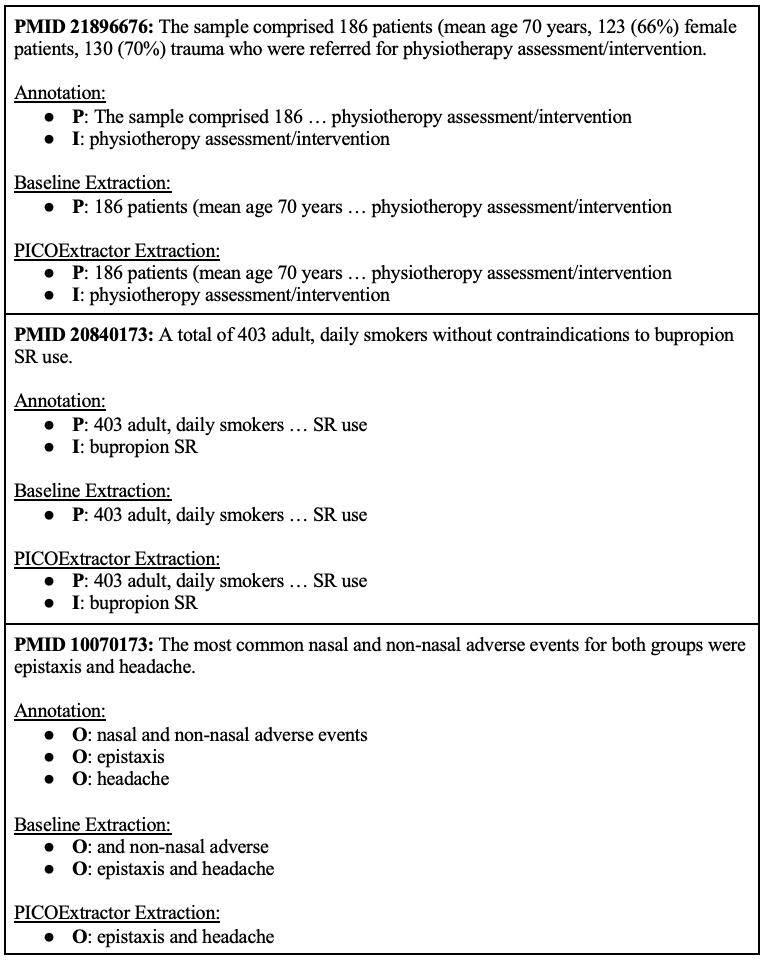}
    \caption{Examples of PICO extraction by PICOX and the baseline.}
    \label{fig:4}
\end{figure}

Figure \ref{fig:4} provides several examples that allow for a detailed comparison of the extraction capabilities of PICOX and the baseline. The first example correctly pointed out the intervention entity ``physiotherapy assessment/intervention'' within the Population entity. Similarly, in the second example, PICOX recognized the Intervention entity ``bupropion SR''. In contrast, the baseline could not extract the Intervention entities that are part of the description of the Population entities. This is because the baseline only assigns a single label to each word, while these Intervention entities should be attributed multiple labels.

We further implemented a data augmentation strategy to enhance the training of our span classifier. The results indicate that the data augmentation strategy effectively reduces false positive errors, leading to higher precision. In addition, even without the data augmentation, our model continues to outperform the baseline on EBM-NLP.

In our approach, we utilized a threshold to determine if a word should be considered as the start or end point of an entity span. As the initial step in our pipeline, our strategy is to include as many plausible span candidates as possible, instead of selectively choosing boundaries with high confidence. This is because false positive spans can be filtered out further down the pipeline. We consider a word as a possible start or end even if the localization model holds uncertainty about its classification. Our results show that PICOX maintains higher precision than the baseline, but the recall falls below the baseline once the threshold exceeds 0.4. The optimal F1 score was achieved when the threshold was within a range from 0.2 and 0.3. However, this optimal threshold value is dataset dependent. In our experiments, we empirically selected a 0.25 threshold for EBM-NLP, PICO-Corpus, and COVID-19 and a 0.4 threshold for AD. We plan to learn these hyperparameters adaptively in the future~\cite{Zhou2021-in}. Other future directions include exploring larger model architectures and distinguishing interventions and comparisons, which were aggregated into one class in most of the existing studies~\cite{Gu2021-gu, Nye2018-jy, Mutinda2022-nm}.

\textbf{\textit{Error analysis and limitations.}} While PICOX has shown advantages in identifying overlapping PICO entities, it has several limitations for further improvement. One lies in accurately locating span boundaries. In the third example in Figure \ref{fig:4}, PICOX overlooked the first Outcome entity, because it did not recognize the starting position, although it correctly recognized the ending position. A more efficient span localization module would alleviate the issue.

The second challenge was accurately identifying boundaries for long spans. In the first example in Figure \ref{fig:4}, the entire sentence was marked as a Population entity. However, PICOX did not include the first three words, leading to the inaccurate determination of span boundaries. In the second example, PICOX identified the entire sentence as the Population entity, even though the annotation does not contain the first three words. 

The third challenge was accurately identifying short spans located near each other, posing difficulty in defining the exact boundaries of an entity. Take, for instance, the third example, where ``epistaxis'' and ``headache'' are annotated as two separate Outcome entities. Our model identified ``epistaxis'' as the beginning of the span and ``headache'' as its end. Consequently, the extracted outcome entity was ``epistaxis and headache'', introducing a false positive error and two false negative errors.

Finally, the span localization and classification modules are trained separately. Moving forward, our plan involves joint training of these two sub-modules. This idea is fueled by the fact that the same encoder structure is shared by the backbones of both these sub-modules. By jointly training them, we can tap into this collective knowledge base for further improvement.

\section{Conclusion}
\label{conclusion}

We introduced a span-based model, PICOX, for recognizing PICO entities from RCT publications. Our model demonstrated improvement in identifying overlapping entities and extracting those of any size without requiring supplementary presumptions to limit the span width. Experimental testing was executed on benchmark datasets across four distinct fields, and our model demonstrated superior results compared to the latest cutting-edge models. A comprehensive examination reveals that span-based models offer valuable perspectives on future pathways for the efficient and effective detection of overlapping PICO elements drawn from RCT publications.

\section*{Supplements}

\setcounter{table}{0}

\begin{table}[!hbpt]
\caption{Performance of PubMedBERT, and PICOX on EBM-NLP, broken down on the length of entities. P - Precision, R - Recall. F1 - F1 score.}\label{app tab:1}
\centering
\footnotesize
\begin{tabular}{lccccccccc}
\toprule
& \multicolumn{3}{c}{\textbf{1 word}} &\multicolumn{3}{c}{\textbf{2-5 word}}& \multicolumn{3}{c}{\textbf{$>$5 word}} \\ 
\cmidrule(lr){2-4} \cmidrule(lr){5-7} \cmidrule(lr){8-10}
& \textbf{P} & \textbf{R} & \textbf{F1} & \textbf{P} & \textbf{R} & \textbf{F1} & \textbf{P} & \textbf{R} & \textbf{F1}\\
\midrule
PubMedBERT & 58.62 & 42.25 & 49.10 & 47.28 & 38.54 & 42.46 & \textbf{45.30} & 45.14 & \textbf{45.22}
\\
PICOX & \textbf{72.41} & \textbf{46.91} & \textbf{56.94} & \textbf{48.64} & \textbf{47.95} & \textbf{48.29} & 35.18 & \textbf{49.76} & 41.22\\
\bottomrule
\end{tabular}
\end{table}

\section*{Funding}

This project was sponsored by the National Library of Medicine grant R01LM009886, R01LM014344, and the National Center for Advancing Clinical and Translational Science award UL1TR001873.



\section*{Conflict of Interest Statement}

The authors declare no competing interests.

\section*{Data Availability}

The data and codes underlying this article will be shared on
\url{https://github.com/WengLab-InformaticsResearch/PICOX}.

\bibliographystyle{unsrtnat}
\bibliography{ref}

\end{document}